\documentclass[aip,pop,preprint,10pt]{revtex4}%
\usepackage{amsfonts}
\usepackage{amsmath}
\usepackage{amssymb}
\usepackage{graphicx}%
\usepackage{color}

\providecommand{\U}[1]{\protect\rule{.1in}{.1in}}
\providecommand{\U}[1]{\protect\rule{.1in}{.1in}}
\providecommand{\U}[1]{\protect\rule{.1in}{.1in}}
\providecommand{\U}[1]{\protect\rule{.1in}{.1in}}

\begin{document}

\title{Hamiltonian magnetohydrodynamics: symmetric formulation, Casimir invariants,  and equilibrium variational principles}

\author{T. Andreussi}
\email{t.andreussi@alta-space.com}
\affiliation{Alta S.p.A., Pisa, 56121, Italy}
\homepage{http://www.alta-space.com}
\author{P. J. Morrison}
\email{morrison@physics.utexas.edu}
\affiliation{Institute for Fusion Studies and Department of Physics, The University of Texas at Austin, Austin, TX 78712-1060, USA}
\author{F. Pegoraro}
\email{pegoraro@df.unipi.it}
\affiliation{Dipartimento di Fisica E.~Fermi, Pisa, 56127, Italy}

\date{\today}

\baselineskip 24 pt

\begin{abstract}

The noncanonical Hamiltonian formulation of magnetohydrodynamics (MHD) is used to construct variational principles for symmetric equilibrium configurations of magnetized plasma including flow. In particular, helical symmetry is considered and results on axial and translational symmetries are retrieved as special cases of the helical configurations. The symmetry condition, which allows the description in terms of  a magnetic flux function, is exploited to deduce a symmetric form of the noncanonical Poisson bracket of MHD.  Casimir invariants are then obtained directly from the Poisson bracket. Equilibria are obtained from an energy-Casimir principle and reduced forms of this variational principle are obtained by the elimination of algebraic constraints.

\end{abstract}
\pacs{52.30.Cv, 02.30.Xx, 47.10.Df, 52.25.Xz}

\keywords{Hamiltonian, Poisson bracket, variational principle, equilibria, symmetry}
\maketitle


\section{Introduction}

Ideal magnetohydrodynamics (MHD) has served as a most important tool for assessing the design and interpretation  of laboratory plasma experiments and for understanding phenomena in naturally occurring plasmas (e.g.\  Refs.~\onlinecite{freidberg,goedbloed}).   Variational principles for equilibria,  or as it is sometimes argued for preferred states,  for a wide variety of geometrical configurations have been discovered over a period of many years.  (e.g.\ Refs.~\onlinecite{woltjer1a,woltjer2a,taylor1,taylor2,hameiri1,goedbloed_evp,ap1,04Ilin}).  In addition, $\delta W$  energy principles  \cite{Bernstein1958b,hain}, which  provide necessary and sufficient conditions for stability of static equilibria,  and other energy-like principles, which provide only sufficient conditions for stability in terms of the Lagrangian displacement variable  \cite{Frieman1960} or in terms of purely Eulerian quantities  \cite{hameiri1,04Ilin},  have been discovered and effectively utilized.

All of the above variational principles for equilibria, which  were for the most part discovered in an {\it ad hoc}  manner,  and all of the energy principles, both Lagrangian and Eulerian,  are a consequence of the fact that  ideal MHD  is a Hamiltonian field theory.   That MHD is Hamiltonian was first shown in terms of the Lagrangian variable description in Ref.~\onlinecite{newcomb} and in terms of the Eulerian variable description in Refs.~\onlinecite{morrison-greene,morrison-greene-add,morrison1} where the noncanonical Poisson bracket was introduced.  In the Hamiltonian context it is seen that  existence of variational principles for equilibrium states is merely the result of the general fact that equilibria are extremal points of Hamiltonian functionals.  Similarly, the existence of the $\delta W$ energy principle for static equilibria is an infinite-dimensional version of Lagrange's stability condition of mechanics (e.g.\ Refs.~\onlinecite{lagrange,koslov}), a consequence of which is that the operator appearing in $\delta W$ is formally self-adjoint because it is a second variation and no further proof is required.  Also, all of the sufficient conditions for stability of equilibria are infinite-dimensional versions of Dirichlet's stability condition \cite{dirichlet,koslov,Morrison1998} and these can be directly derived from the Hamiltonian formulation.   (For discussion of these ideas in the ideal fluid context see  Ref.~\onlinecite{Morrison1998}.)

The purpose of the present paper and its companion  \cite{amp2} is to continue with the approach of  Ref.~\onlinecite{amp0}, which starts from  the noncanonical Poisson bracket of Refs.~\onlinecite{morrison-greene,morrison-greene-add,morrison1} and then reduces to obtain the Hamiltonian formulations for translational and rotational symmetry.  Here we generalize and obtain an inclusive  Hamiltonian description for any metric symmetry.  From the noncanonical Poisson bracket we derive large families of Casimir invariants that are then used to obtain general variational principles for equilibria, including equilibria with helical symmetry and flow.  This  prepares the way for our companion paper  \cite{amp2},  where we consider stability via several approaches.

Specifically, in Sec.~\ref{review} we briefly review the Hamiltonian description of MHD as given in Refs.~\onlinecite{morrison-greene,morrison-greene-add,morrison1}.  This is followed in Sec.~\ref{reduction} by the symmetry reduction, which is done by effecting the chain rule for functional derivatives.  Then in Sec.~\ref{casimirs},  Casimir invariants are obtained directly from the noncanonical Poisson bracket, and this allows us to construct  the equilibrium variational principles in Sec.~\ref{vps}.   These variational principles are then reduced by the elimination of algebraic constraints to obtain variation principles for special cases.  In Sec.~\ref{conclu} several applications of helical equilibria, both with and without flow are discussed.  

\section{Noncanonical Hamiltonian description of MHD}
\label{review}

Following Morrison and Greene \cite{morrison-greene}, the ideal dynamics of MHD plasma is described
in terms of the Eulerian variables $Z:=\left(  \rho,\mathbf{v},s,\mathbf{B}\right)$, i.e.,  the plasma density $\rho$, the flow velocity
$\mathbf{v}$, the magnetic field $\mathbf{B}$\ and the entropy per unit
mass,  $s$ (or alternatively the plasma temperature
or pressure),  in  the Hamiltonian  form%
\begin{equation}
\frac{\partial Z}{\partial t}=\left\{Z,H\right\}_Z
\,,\label{eq:genEOM}%
\end{equation}
where $H$\ is  the Hamiltonian for MHD corresponding to the energy,%
\begin{equation}
H=\int_{V}\left[  \frac{1}{2}\rho v^{2}+\rho U+\frac{1}{8\pi}B^{2}\right]
d^{3}r\,,\label{eq:H}%
\end{equation}
and $\left\{  \cdot,\cdot\right\}  $ represents the noncanonical Poisson
bracket of MHD. In Eq.~(\ref{eq:H}) the function $U=U\left(  \rho,s\right)$
represents the internal energy of the plasma, which is related to the plasma
pressure and temperature by the relationships $p=\rho^{2}{\partial U}/{\partial\rho}$ and $T={\partial U}/{\partial
s}$;  we note that gravitational effects could be included by adding a term $\rho \varphi$ to the integrand where $\varphi$ is an external potential.  The bracket of Eq.~(\ref{eq:genEOM}), which follows from the canonical
Hamiltonian formulation of Newcomb  \cite{newcomb} through the transformation from canonical
Lagrangian to noncanonical Eulerian variables, is given by%
\begin{align}
\left\{  F,G\right\}_Z   &  =-%
{\displaystyle\int\nolimits_{V}}
\bigg\{  F_{\rho}\mathbf{\nabla}\cdot G_{\mathbf{v}}-G_{\rho}\mathbf{\nabla
}\cdot F_{\mathbf{v}}  \nonumber\\
&  +\frac{\mathbf{\nabla\times v}}{\rho}\cdot\left(  G_{\mathbf{v}}\times
F_{\mathbf{v}}\right)  \nonumber\\
&  +\frac{\mathbf{\nabla}s}{\rho}\cdot\left(  F_{s}G_{\mathbf{v}}%
-G_{s}F_{\mathbf{v}}\right)  \nonumber\\
&  +\mathbf{B}\cdot\left[  \left(  \frac{1}{\rho}F_{\mathbf{v}}\cdot
\mathbf{\nabla}\right)  G_{\mathbf{B}}-\left(  \frac{1}{\rho}G_{\mathbf{v}%
}\cdot\mathbf{\nabla}\right)  F_{\mathbf{B}}\right]  \nonumber\\
&     +\mathbf{B}\cdot\left[  \left(  \mathbf{\nabla}\frac{1}{\rho
}F_{\mathbf{v}}\right)  \cdot G_{\mathbf{B}}-\left(  \mathbf{\nabla}\frac
{1}{\rho}G_{\mathbf{v}}\right)  \cdot F_{\mathbf{B}}\right]  \bigg\}
d^{3}r,\label{eq:bra0}%
\end{align}
where $F$ and $G$ are two generic functionals and subscripts indicate
functional derivatives.

Given a generic functional $F$, the functional derivative is defined by $
\delta F=\int_V F_{Z}\cdot\delta Z\,  d^{3}r$ (cf., e.g., \cite{Morrison1998})
and, in particular, the functional derivatives of the Hamiltonian (\ref{eq:H}) with respect to the variables $Z$ are%
\begin{equation}
H_{\rho}=\frac{1}{2}v^{2}+U+\frac{p}{\rho},\qquad H_{\mathbf{v}}%
=\rho\mathbf{v},\mathbf{\qquad}H_{s}=\rho T,\qquad H_{\mathbf{B}}=\frac
{1}{4\pi}\mathbf{B}.\label{eq:H_der}%
\end{equation}
The functional derivatives of the variables $Z$  can be calculated by making use
of the identity%
\begin{equation}
Z\left(\mathbf{x} \right)=\int_V  Z\left(\mathbf{x}^{\prime} \right)\, \delta\left(  \mathbf{x}^{\prime}-\mathbf{x}%
\right)  \,  d^{3}r\,,
\label{eq:dlZ}
\end{equation}
giving, for example, $\delta \rho(\mathbf{x})/\delta \rho(\mathbf{x}^{\prime})=\delta(\mathbf{x}^{\prime}-\mathbf{x})$, which removes the integral of the Poisson bracket when evaluating (\ref{eq:genEOM}).

Substituting   expressions (\ref{eq:H_der}) and (\ref{eq:dlZ}) into Eqs.~(\ref{eq:genEOM}), we obtain the equations of MHD,
\begin{align}
\frac{\partial\rho}{\partial t} &  =-\mathbf{\nabla}\cdot\left(
\rho\mathbf{v}\right)  ,\label{eq:mass}\\
\frac{\partial\mathbf{v}}{\partial t} &  =-\mathbf{\nabla}\left(  \frac{v^{2}%
}{2}+U+\frac{p}{\rho}\right)  -\left(  \mathbf{\nabla\times v}\right)
\times\mathbf{v}+T\mathbf{\nabla}s+\frac{1}{4\pi\rho}\left(  \mathbf{\nabla
}\times\mathbf{B}\right)  \times\mathbf{B},\label{eq:mome}\\
\frac{\partial s}{\partial t} &  =-\mathbf{v}\cdot\mathbf{\nabla
}s,\label{eq:entr}\\
\frac{\partial\mathbf{B}}{\partial t} &  =-\mathbf{\nabla}\times\left(
\mathbf{B}\times\mathbf{v}\right)  ,\label{eq:fara}%
\end{align}
where Eq.~(\ref{eq:mass}) represents  mass conservation equation, Eq.~(\ref{eq:mome}) momentum balance, Eq.~(\ref{eq:entr})  entropy  advection,  and Eq.~(\ref{eq:fara}) is Faraday's law for a perfectly conductive medium. In actuality, the Poisson bracket of (\ref{eq:bra0}) gives MHD in conservation form, in which    Eqs.~(\ref{eq:mome}) and (\ref{eq:fara}) differ by terms involving $\nabla\cdot \mathbf{B}$, but this will not bear on our development.  (In Ref.~\onlinecite{morrison1} it was shown that $\nabla\cdot \mathbf{B}=0$ is not needed for MHD to be Hamiltonian and the results of Ref.~\onlinecite{gudunov} indicate that  the conservation form is superior for numerical computation.)

The Poisson bracket of (\ref{eq:bra0}) can be rewritten in terms of any complete set of  variables -- switching from one set to another amounts to a change of coordinates.   A convenient form of the MHD Poisson bracket is obtained by using, instead of
the variables $\mathbf{v}$ and $s$, the density variables $\mathbf{M}%
=\rho\mathbf{v}$ and $\sigma=\rho s$. We let $\mathcal{Z}:=(\rho, \mathbf{M}, \sigma, \mathbf{B})$ denote the new set.   To transform from $Z$ to $\mathcal{Z}$  we use the functional  chain rule identities,  %
\begin{equation}
\left.  F_{\rho}\right\vert _{\mathbf{v},s}=\left.  F_{\rho}\right\vert
_{\mathbf{M},\sigma}+\mathbf{v}\cdot F_{\mathbf{M}}+sF_{\sigma},\qquad
F_{\mathbf{v}}=\rho F_{\mathbf{M}},\qquad F_{s}=\rho F_{\sigma}\,,
\end{equation}
with $F_{\mathbf{B}}$ unchanged,  to transform the Poisson bracket of (\ref{eq:bra0}) into%
\begin{align}
\left\{  F,G\right\}_{\mathcal{Z}}   &  =-%
{\displaystyle\int\nolimits_{V}}
\Big\{  \rho\left(  F_{\mathbf{M}}\cdot\mathbf{\nabla}G_{\rho}-G_{\mathbf{M}%
}\cdot\mathbf{\nabla}F_{\rho}\right)    \nonumber\\
&  +\mathbf{M}\cdot\left[  \left(  F_{\mathbf{M}}\cdot\mathbf{\nabla}\right)
G_{\mathbf{M}}-\left(  G_{\mathbf{M}}\cdot\mathbf{\nabla}\right)
F_{\mathbf{M}}\right] \nonumber\\
&  +\sigma\left(  F_{\mathbf{M}}\cdot\mathbf{\nabla}G_{\sigma}-G_{\mathbf{M}%
}\cdot\mathbf{\nabla}F_{\sigma}\right) \nonumber\\
&  +\mathbf{B}\cdot\big[  \left(  F_{\mathbf{M}}\cdot\mathbf{\nabla}\right)
G_{\mathbf{B}}-\left(  G_{\mathbf{M}}\cdot\mathbf{\nabla}\right)
F_{\mathbf{B}}\big] \nonumber\\
&    +\mathbf{B}\cdot\left(  \mathbf{\nabla}F_{\mathbf{M}}\cdot
G_{\mathbf{B}}-\mathbf{\nabla}G_{\mathbf{M}}\cdot F_{\mathbf{B}}\right)
\Big\} d^{3}r\,. \label{eq:bra}%
\end{align}
The bracket of (\ref{eq:bra}) is the  Lie-Poisson bracket (see Ref.~\onlinecite{Morrison1998}), i.e., a  bracket linear in each variable, obtained in  Ref.~\onlinecite{morrison-greene}.


\section{Symmetric MHD}
\label{reduction}

All geometric symmetries can be described as a combination of axial and translational symmetry, a breakdown of  helical
symmetry.  Given a cylindrical coordinate system $\left(r,\phi,z\right)$,
we define a helical coordinate $u=\phi\left[  l\right]  \sin\alpha+z\cos\alpha$,
where $\left[l\right]$  is a scale length and $\alpha$ defines the
helical angle. The unit vector in the direction of  the coordinate $u$ can
be written as%
\begin{equation}
\mathbf{u}=kr\nabla u=   \hat{\mathbf{\phi}} \,  k\left[  l\right]  \sin\alpha
+ \hat{\mathbf{z}}\,   kr\cos\alpha\, ,
\end{equation}
where $k=\left(\left[  l\right]  ^{2}\sin^{2}\alpha+r^{2}\cos^{2}%
\alpha\right)^{-1/2}$ represents a metric factor. The second helical
direction is given in terms of the following unit vector:
\begin{equation}
\mathbf{h=}kr\mathbf{\nabla}r\times\mathbf{\nabla}u=-   \hat{\mathbf{\phi}}\, kr\cos
\alpha+\hat{\mathbf{z}}\,  k\left[l\right]  \sin\alpha\, ,
\end{equation}
and the helical symmetry is expressed by the fact that $\mathbf{h}\cdot\mathbf{\nabla}f=0$,
where $f$ is a generic scalar function. The direction $\mathbf{h}$,
called the symmetry direction, can be chosen to obtain axial ($\alpha=0$),
translational ($\alpha=\pi/2$),  or {true} helical ($0<\alpha<\pi/2$)
symmetry, with the metric factor $k$ changing accordingly.  In the
following,  we use the identities, %
\begin{equation}
\mathbf{\nabla}\cdot\mathbf{h}=0\qquad{\rm and}\qquad \mathbf{\nabla}\times\left(
k\mathbf{h}\right)  =-  \mathbf{h}\,  k^{3}\left[ l\right]  \sin2\alpha \,,
\end{equation}
which imply for $\sin2\alpha=0$  the existence of the
coordinate $\mathbf{\nabla}h=k\mathbf{h}$ in the symmetry direction.

Using the notation described before, the magnetic field and the mass flow can be rewritten as%
\begin{equation}
\mathbf{B}\left(  r,u\right)  =B_{h}\left(  r,u\right)  \mathbf{h}%
+\mathbf{B}_{\bot}\left(  r,u\right)\,,  \qquad
\mathbf{M}\left(  r,u\right)  =M_{h}\left(  r,u\right)  \mathbf{h}%
+\mathbf{M}_{\bot}\left(  r,u\right)  \label{eq:Mhel}%
\end{equation}
and, since $ \mathbf{\nabla}\cdot\mathbf{B}=0$, the magnetic field perpendicular to the symmetry direction can be expressed in terms of a magnetic flux function  $\psi=\psi\left(  r,u\right)$ as $\mathbf{B}_{\bot}\left(  r,u\right)=\mathbf{\nabla}\psi\times k\mathbf{h}$.

Given a generic functional $F$ and using the chain rule, the following functional
derivative relations result:%
\begin{equation}
F_{B_{h}}=F_{\mathbf{B}}\cdot\mathbf{h}\,, \qquad F_{\psi}=\mathbf{\nabla}
\cdot\left(  F_{\mathbf{B}}\times k\mathbf{h}\right)\,, \qquad {\rm and}\qquad
F_{\mathbf{M}}=F_{M_{h}}\mathbf{h}+F_{\mathbf{M}_{\bot}} \label{eq:fm}%
 \,.
\end{equation}
In term of the variables $\mathcal{Z}_{S}:=\left(  \rho,\mathbf{M}_{\bot},M_{h}%
,\sigma,\psi,B_{h}\right)$ the Poisson
bracket of Eq.~(\ref{eq:bra0}) transforms into  the  ``symmetric" MHD bracket given by
\begin{align}
\left\{  F,G\right\}  _{\mathcal{SYM}}  &  =-%
{\displaystyle\int\nolimits_{V}}
\Big\{  \rho\left(  F_{\mathbf{M}_{\bot}}\cdot\mathbf{\nabla}G_{\rho
}-G_{\mathbf{M}_{\bot}}\cdot\mathbf{\nabla}F_{\rho}\right)
\nonumber\\
&  +M_{h}\big[  F_{\mathbf{M}_{\bot}}\cdot\mathbf{\nabla}\left(
kG_{M_{h}}\right)  -G_{\mathbf{M}_{\bot}}\cdot\mathbf{\nabla}\left(
kF_{M_{h}}\right)  \big] /k \nonumber\\
&  +  \left( k^{2}\left[  l\right]  \sin2\alpha\right)   M_{h} \, \mathbf{h}%
\cdot\left(  F_{\mathbf{M}_{\bot}}\times G_{\mathbf{M}_{\bot}}\right)
\nonumber\\
&  +\mathbf{M}_{\bot}\cdot\big[  \left(  F_{\mathbf{M}_{\bot}}\cdot
\mathbf{\nabla}\right)  G_{\mathbf{M}_{\bot}}-\left(  G_{\mathbf{M}_{\bot}%
}\cdot\mathbf{\nabla}\right)  F_{\mathbf{M}_{\bot}}\big] \nonumber\\
&  +\sigma\left(  F_{\mathbf{M}_{\bot}}\cdot\mathbf{\nabla}G_{\sigma
}-G_{\mathbf{M}_{\bot}}\cdot\mathbf{\nabla}F_{\sigma}\right) \nonumber\\
&  +kB_{h}\big[  F_{\mathbf{M}_{\bot}}\cdot\mathbf{\nabla}\left(
G_{B_{h}}/k\right)  -G_{\mathbf{M}_{\bot}}\cdot\mathbf{\nabla}\left(
 F_{B_{h}}/k \right)  \big] \nonumber\\
&  +\psi\left(  F_{\mathbf{M}_{\bot}}\cdot\mathbf{\nabla}G_{\psi
}-G_{\mathbf{M}_{\bot}}\cdot\mathbf{\nabla}F_{\psi}\right)  -\psi\left(
F_{\psi}\mathbf{\nabla}\cdot G_{\mathbf{M}_{\bot}}-G_{\psi}\mathbf{\nabla
}\cdot F_{\mathbf{M}_{\bot}}\right) \nonumber\\
&  -\left(  k^{3}\left[  l\right]  \sin2\alpha\right)  \mathbf{\nabla}%
\psi\cdot\left(  F_{B_{h}}G_{\mathbf{M}_{\bot}}-G_{B_{h}}F_{\mathbf{M}_{\bot}%
}\right) \nonumber\\
&    + \psi\big(  \left[ G_{B_{h}}/k ,  kF_{M_{h}%
}\right]  -\left[ F_{B_{h}}/k , kG_{M_{h}}\right]  \big)
\Big\}  d^{3}r\,. \label{eq:bra_sym}%
\end{align}
where $\left[  F,G\right]  :=\left(  \mathbf{\nabla}F\times\mathbf{\nabla}G\right)
\cdot k\mathbf{h}$.  Because this calculation is similar to one of Ref.~\onlinecite{amp0},  we forgo the details.

Using (\ref{eq:bra_sym}) the  equations for symmetric MHD dynamics are obtained
\begin{align}
& \frac{\partial\rho}{\partial t} =-\mathbf{\nabla}\cdot\mathbf{M}_{\bot},\label{eq:dyn1}\\
& \frac{\partial M_{h}}{\partial t} =-k\mathbf{\nabla}\cdot\left(
\mathbf{M}_{\bot}\frac{M_{h}}{k\rho}\right)  +k\left[  \psi\mathbf{,}\frac
{1}{4\pi k}B_{h}\right] , \\
& \frac{\partial\mathbf{M}_{\bot}}{\partial t} =-\rho\mathbf{\nabla}\left(
\frac{M^{2}}{2\rho^{2}}+U+\frac{p}{\rho}\right)  -\left(  \mathbf{\nabla
}\times\frac{\mathbf{M}_{\bot}}{\rho}\right)  \times\mathbf{M}_{\bot}-\left(
\mathbf{\nabla}\cdot\mathbf{M}_{\bot}\right)  \frac{\mathbf{M}_{\bot}}{\rho
}\nonumber\\
&\qquad  +kM_{h}\mathbf{\nabla}\frac{M_{h}}{k\rho}+\left(  k^{2}\left[
l\right]  \sin2\alpha\right)  \left(  \frac{M_{h}}{\rho}\mathbf{h}%
\times\mathbf{M}_{\bot}\right)  +\rho T\mathbf{\nabla}\frac{\sigma}{\rho}-\mathbf{\nabla}\cdot\left(
\frac{k^{2}}{4\pi}\mathbf{\nabla}\psi\right)  \mathbf{\nabla}\psi\nonumber\\
&\qquad -kB_{h}\mathbf{\nabla}\frac{B_{h}}{4\pi k}-\left(
k^{3}\left[  l\right]  \sin2\alpha\right)  \frac{B_{h}}{4\pi}\mathbf{\nabla
}\psi,\\
& \frac{\partial\sigma}{\partial t} =-\mathbf{\nabla}\cdot\left(
\mathbf{M}_{\bot}\frac{\sigma}{\rho}\right)  ,\\
& \frac{\partial B_{h}}{\partial t} =-\frac{1}{k}\mathbf{\nabla}\cdot\left(
\mathbf{M}_{\bot}\frac{kB_{h}}{\rho}\right)  +\left(  k^{3}\left[  l\right]
\sin2\alpha\right)  \mathbf{\nabla}\psi\cdot\frac{\mathbf{M}_{\bot}}{\rho
}+\frac{1}{k}\left[  \psi\mathbf{,}\frac{kM_{h}}{\rho}\right]  ,\\
& \frac{\partial\psi}{\partial t} =-\mathbf{\nabla}\psi\cdot\frac
{\mathbf{M}_{\bot}}{\rho}.\label{eq:dyn6}
\end{align}
In comparison  to Eqs.~(\ref{eq:mass})--(\ref{eq:fara}), the number of equations needed to describe the symmetric dynamics is reduced because of the introduction of  $\psi$.  Moreover, the differential operator $\mathbf{\nabla}$ in Eqs.~(\ref{eq:dyn1})--(\ref{eq:dyn6})   only depends on  $u$ and $r$.


\section{Symmetric Casimirs}
\label{casimirs}

Now we seek the Casimir invariants associated with the helically-symmetric MHD
bracket (\ref{eq:bra_sym}), i.e.\ functionals $C$ that satisfy $\left\{  F,C\right\}_{\mathcal{SYM}}=0$
for all functionals $F$.  With (\ref{eq:bra_sym}) we see that $\left\{  F,C\right\}_{\mathcal{SYM}}=0$  implies
\begin{equation}
\int_{V}\left[  F_{\rho}\mathfrak{C}_{1}+kF_{M_{h}}\mathfrak{C}_{2}+F_{\sigma
}\mathfrak{C}_{3}+\frac{1}{k}F_{B_{h}}\mathfrak{C}_{4}+F_{\psi}\mathfrak{C}%
_{5}+F_{\mathbf{M}_{\bot}}\cdot\mathfrak{C}_{6}\right]  d^{3}r=0,
\end{equation}
where the functions $\mathfrak{C}_{i}$ are given by
\begin{align}
\mathfrak{C}_{1}  &  =-\mathbf{\nabla}\cdot\left(  \rho C_{\mathbf{M}_{\bot}%
}\right)  ,\label{eq:cas0_1}\\
\mathfrak{C}_{2}  &  =-\mathbf{\nabla}\cdot\left(  \frac{1}{k}M_{h}%
C_{\mathbf{M}_{\bot}}\right)  -\left[  \psi,\frac{1}{k}C_{B_{h}}\right]
,\label{eq:cas0_2}\\
\mathfrak{C}_{3}  &  =-\mathbf{\nabla}\cdot\left(  \sigma C_{\mathbf{M}_{\bot
}}\right)  ,\label{eq:cas0_3}\\
\mathfrak{C}_{4}  &  =-\mathbf{\nabla}\cdot\left(  kB_{h}C_{\mathbf{M}_{\bot}%
}\right)  -\left[  \psi,kC_{M_{h}}\right]  +\left(  k^{4}\left[  l\right]
\sin2\alpha\right)  \left(  \mathbf{\nabla}\psi\cdot C_{\mathbf{M}_{\bot}%
}\right)  ,\label{eq:cas0_4}\\
\mathfrak{C}_{5}  &  =-\mathbf{\nabla}\psi\cdot C_{\mathbf{M}_{\bot}%
},\label{eq:cas0_5}\\
\mathfrak{C}_{6}  &  =-\rho\mathbf{\nabla}C_{\rho}-\frac{M_{h}}{k}%
\mathbf{\nabla}\left(  kC_{M_{h}}\right)  -\left(  k^{2}\left[  l\right]
\sin2\alpha\right)  M_{h}\left(  C_{\mathbf{M}_{\bot}}\times\mathbf{h}\right)
+\label{eq:cas0_9}\\
&  -\left[  \left(  \mathbf{\nabla}\times\mathbf{M}_{\bot}\right)  \times
C_{\mathbf{M}_{\bot}}+\mathbf{\nabla}\left(  \mathbf{M}_{\bot}\cdot
C_{\mathbf{M}_{\bot}}\right)  +\left(  \mathbf{\nabla}\cdot C_{\mathbf{M}%
_{\bot}}\right)  \mathbf{M}_{\bot}\right]  +\nonumber\\
&  -\sigma\mathbf{\nabla}C_{\sigma}-kB_{h}\mathbf{\nabla}\left(  \frac{1}%
{k}C_{B_{h}}\right)  +C_{\psi}\mathbf{\nabla}\psi-\left(  k^{3}\left[
l\right]  \sin2\alpha\right)  C_{B_{h}}\mathbf{\nabla}\psi\nonumber\,.
\end{align}
Since each term in the bracket must vanish separately,
this implies the Casimir conditions $\mathfrak{C}_{i}=0$ for $\ i=1\ldots6$.

We first investigate the case where $C_{\mathbf{M}_{\bot}}=0$, which implies the reduced set of conditions
\begin{align}
\mathfrak{C}_{2}  &  =-\left[  \psi,\frac{1}{k}C_{B_{h}}\right]
,\label{eq:cas0_2red}\\
\mathfrak{C}_{4}  &  =-\left[  \psi,kC_{M_{h}}\right]  ,\label{eq:cas0_4red}\\
\mathfrak{C}_{6}  &  =-\rho\mathbf{\nabla}C_{\rho}-\frac{M_{h}}{k}%
\mathbf{\nabla}\left(  kC_{M_{h}}\right)  +\label{eq:cas0_9red}\\
&  -\sigma\mathbf{\nabla}C_{\sigma}-kB_{h}\mathbf{\nabla}\left(  \frac{1}%
{k}C_{B_{h}}\right)  +C_{\psi}\mathbf{\nabla}\psi-\left(  k^{3}\left[
l\right]  \sin2\alpha\right)  C_{B_{h}}\mathbf{\nabla}\psi\nonumber\,.
\end{align}

Upon substituting the functional
\begin{equation}
C_{1}=\int_{V}\rho\mathcal{J}\left(  \frac{\sigma}{\rho},\psi,\frac{1}{\rho
}\left[  \frac{\sigma}{\rho},\psi\right]  ,\frac{1}{\rho}\left[  \frac{1}%
{\rho}\left[  \frac{\sigma}{\rho},\psi\right]  ,\psi\right]  ,\frac{1}{\rho
}\left[  \frac{\sigma}{\rho},\frac{1}{\rho}\left[  \frac{\sigma}{\rho}%
,\psi\right]  \right]  ,...\right)  d^{3}r \label{eq:C1}%
\end{equation}
into Eqs.~(\ref{eq:cas0_2red})--(\ref{eq:cas0_9red}), it is
straightforward to prove that $C_{1}$\ is a Casimir. In fact, since
\begin{equation}
\frac{1}{\rho}\left[  \frac{\sigma}{\rho},\psi\right]  =\frac{\mathbf{B}}{\rho
}\cdot\mathbf{\nabla}\frac{\sigma}{\rho} \qquad {\rm and}\qquad
\frac{1}{\rho}\left[  \frac{1}{\rho}\left[  \frac{\sigma}{\rho},\psi\right]
,\psi\right]    =\frac{\mathbf{B}}{\rho} \cdot\nabla\left( \frac{\mathbf{B}}{\rho
}\cdot\nabla \frac{\sigma}{\rho}\right)
\end{equation}
the Casimir (\ref{eq:C1}) is similar but not equivalent to  one of Refs.~\onlinecite{padhye1} and \onlinecite{padhye2}, which is more general than the
one described in Ref.~\onlinecite{amp0}.   This Casimir is akin to Ertel's theorem of geophysical fluid dynamics.

Next, from   conditions (\ref{eq:cas0_2red})
and (\ref{eq:cas0_9red}) we deduce that%
\begin{equation}
C_{2}=\int_{V}\left[  kB_{h}\mathcal{H}\left(  \psi\right)  +\left(
k^{4}\left[  l\right]  \sin2\alpha\right)    {\mathcal{H}^{-}(\psi)}   \right]  d^{3}r\,,
 \label{eq:C2}%
\end{equation}
where $\mathcal{H}^{-}(\psi):=\int^{\psi}\mathcal{H}\left(  \psi^{\prime}\right)\,
d\psi^{\prime}$, is also a Casimir and, analogously, from   condition (\ref{eq:cas0_4red}) we
obtain the Casimir%
\begin{equation}
C_{3}=\int_{V}\frac{1}{k}M_{h}\mathcal{G}\left(  \psi\right)  d^{3}r.
\label{eq:C3}%
\end{equation}

If we suppose $C_{\mathbf{M}_{\bot}}\neq0$, then from condition
(\ref{eq:cas0_5}), it follows that
\begin{equation}
C_{\mathbf{M}_{\bot}}=\mathbf{\nabla}\psi\times Ak\mathbf{h}
\label{eq:C_Mperp}%
\end{equation}
where $A$ is a generic function. Thus, we can rewrite the conditions
(\ref{eq:cas0_1})--(\ref{eq:cas0_4}) as%
\begin{align}
\mathfrak{C}_{1} &  =\left[  \psi,\rho A\right]  ,\label{eq:cas0_1-1}\\
\mathfrak{C}_{2} &  =\left[  \psi,\frac{M_{h}}{k}A-\frac{1}{k}C_{B_{h}%
}\right]  ,\label{eq:cas0_2-1}\\
\mathfrak{C}_{3} &  =\rho A\left[  \psi,\frac{\sigma}{\rho}\right]
,\label{eq:cas0_3-1}\\
\mathfrak{C}_{4} &  =\left[  \psi,kB_{h}A-kC_{M_{h}}\right]
,\label{eq:cas0_4-1}%
\end{align}
which  implies that, unless (see Eq.~(\ref{eq:cas0_3-1}))%
\begin{equation}
\left[  \psi,\frac{\sigma}{\rho}\right]  =0\,,
\label{eq:cond}%
\end{equation}
no further Casimir functionals can be found. It can be easily shown that
condition (\ref{eq:cond}) holds for stationary flows and vice versa
(from $\mathbf{\nabla}\cdot \mathbf{M} =0$, we deduce
$\mathbf{M}=\mathbf{\nabla}\chi\times k\mathbf{h}$ and using the
perfect conductivity equation we obtain $\left[  \psi,\chi\right]  =0$.
Analogously, the entropy equation becomes $\left[  \sigma/ \rho,\chi\right]
=0$ and, except where $\mathbf{\nabla}\chi=0$  and
$\mathbf{\nabla}\psi\neq0$, $\left[ \sigma/\rho,\psi\right]  =0$). If Eq.~(\ref{eq:cond}) holds, from condition
(\ref{eq:cas0_1-1}) we obtain $A={\mathcal{F}}/{\rho}$,  where $\mathcal{F}$\ is a generic function of $\psi$\ or $\sigma/\rho$, and
conditions (\ref{eq:cas0_2-1}) and (\ref{eq:cas0_4-1}) imply%
\begin{equation}
C_{B_{h}}=\frac{M_{h}}{\rho}\mathcal{F}\qquad{\rm and}\qquad  C_{M_{h}}=\frac{B_{h}}{\rho
}\mathcal{F},\label{eq:condA2}%
\end{equation}
plus solutions in the form of (\ref{eq:C2}) and (\ref{eq:C3}). By integrating
conditions (\ref{eq:C_Mperp}) and (\ref{eq:condA2}), we obtain%
\begin{equation}
C_{4}=\int_{V}\left(  \mathbf{M}_{\bot}\cdot
\mathbf{B}_{\bot}+M_{h}B_{h}\right) {\mathcal{F}}/{\rho}\,  d^{3}r=\int_{V}
\mathbf{v\cdot B}\, \mathcal{F} \,  d^{3}r,\label{eq:C4}%
\end{equation}
which also satisfies  condition (\ref{eq:cas0_9}), and is thus  a Casimir.

For flows that satisfy condition (\ref{eq:cond}), the family of invariants
(\ref{eq:C1}) can be rewritten in the simpler form%
\begin{equation}
C_{1}=\int_{V}\rho\mathcal{J}d^{3}r, \label{eq:C1_bis}%
\end{equation}
where $\mathcal{J}$\ is a generic function of $\psi$\ or $\sigma/\rho$.

Since Casimirs are conserved quantities, their integrands, say  $\mathcal{C}_i$,  are  densities associated with the `currents' $\mathbf{J}_i$ that satisfy  conservation equations of the form $ {\partial \mathcal{C}_i }/{\partial t}+ \nabla\cdot \mathbf{J}_i=0$,
where  $i=1\dots 4$.  These Casimir currents are given by
\begin{eqnarray}
 \mathbf{J}_1&=&  \mathbf{M}_{\bot} \mathcal{J}
 \nonumber\\
 \mathbf{J}_2&=&\left(  \mathbf{M}_{\bot}%
\frac{kB_{h}}{\rho}+\frac{kM_{h}}{\rho}\mathbf{B}_{\bot}\right)    \mathcal{H}
  \nonumber\\
 \mathbf{J}_3&=&\left(  \mathbf{M}_{\bot}%
\frac{M_{h}}{k\rho}+\frac{B_{h}}{4\pi k}\mathbf{B}_{\bot}\right)   \mathcal{G}
  \nonumber\\
 \mathbf{J}_4&=&   \mathbf{M}\times\left(
\mathbf{B}\times\mathbf{M}\right)  \frac{\mathcal{F}}{\rho^{2}}  -\mathbf{B}\left(\frac{M^{2}}{2\rho^{2}}+U + \frac{p}{\rho} 
\right) \mathcal{F}
 \,.
 \end{eqnarray}
If we assume  the bounding surface is a {fixed magnetic surface, i.e. $\mathbf{n}\cdot\mathbf{B}=0$  and $\mathbf{n}\cdot \mathbf{M}=0$,   this surface respects the symmetry, and the unit surface normal $\mathbf{n}$  satisfies $\mathbf{n}\cdot \mathbf{h}=0$.  Consequently, $\mathbf{n}\cdot\mathbf{B}_{\bot}=0$  and $\mathbf{n}\cdot \mathbf{M}_{\bot}=0$}. 
Thus, for this kind of fixed boundary condition,  the Casimirs are conserved.   However, the possibility of Casimir injection exists and in a future publication we will consider more general boundary conditions.


\section{Variational principle and equilibria}
\label{vps}

Now we proceed to construct the energy-Casimir variational principle for
symmetric MHD equilibria. With the knowledge that extrema of the energy-Casimir
functional must correspond to equilibria, we consider
\begin{equation}
\mathfrak{F}=H-\int_{V}\rho\mathcal{J}\, d^{3}r-\int_{V}\left[  kB_{h}%
\mathcal{H}+\left(  k^{4}\left[  l\right]  \sin2\alpha\right)
 \mathcal{H}^{-}   \right]  d^{3}r-\int_{V}\frac{1}{k}%
M_{h}\mathcal{G}\, d^{3}r-\int_{V}\mathbf{v\cdot B}\,\mathcal{F}d^{3}r,
\label{eq:enerCas}%
\end{equation}
where the Hamiltonian (\ref{eq:H}) is expressed in terms of symmetric variables $\mathcal{Z}_S$ as%
\begin{equation}
H=%
{\displaystyle\int_{V}}
\left(  \frac{M_{h}^{2}}{2\rho}+\frac{M_{\bot}^{2}}{2\rho}+\rho U+\frac
{k^{2}\left\vert \nabla\psi\right\vert ^{2}}{8\pi}+\frac{B_{h}^{2}}{8\pi
}\right)  d^{3}r
\end{equation}
and $\mathcal{F}$,$\ \mathcal{G}$, $\mathcal{H}$ ($\mathcal{H}^{-}$), and $\mathcal{J}$ are four
arbitrary functions of $\psi$. Moreover, in order to satisfy Eq.~(\ref{eq:cond}) we consider $\sigma/\rho=\mathcal{S}\left(  \psi\right)  $.
Thus, the constrained energy in terms of the variables $\mathcal{Z}_{S}$ is given by %
\begin{align}
\mathfrak{F}[ \mathcal{Z}_{S}] &  =%
{\displaystyle\int_{V}}
\left(  \frac{M_{\bot}^{2}}{2\rho}+\frac{M_{h}^{2}}{2\rho}+\rho U+\frac
{k^{2}\left\vert \nabla\psi\right\vert ^{2}}{8\pi}+\frac{B_{h}^{2}}{8\pi
}  -  \rho\mathcal{J} \right.
\nonumber\\
&  \left. \hspace{1.0 cm} - kB_{h}\mathcal{H}-  k^{4}\left[  l\right]
\sin2\alpha\,    \mathcal{H}^{-} -\frac{1}{k}%
M_{h}\mathcal{G}-\frac{\mathbf{M}}{\rho}\cdot\mathbf{B}\,\mathcal{F}\right)
d^{3}r,
\end{align}
or in terms of the variables $Z_{S}:=\left(  \rho,\mathbf{v}_{\bot
},v_{h},\psi,B_{h}\right)$ is given by
\begin{align}
\mathfrak{F}  [Z_S] &  =%
{\displaystyle\int_{V}}
\left(  \frac{\rho v_{\bot}^{2}}{2}+\frac{\rho v_{h}^{2}}{2}+\rho
U+\frac{k^{2}\left\vert \nabla\psi\right\vert ^{2}}{8\pi}+\frac{B_{h}^{2}%
}{8\pi}\right.  - \rho\mathcal{J}\nonumber\\
&  \left.  \hspace{1.0 cm} -kB_{h}\mathcal{H}-   k^{4}\left[  l\right]
\sin2\alpha \,    \mathcal{H}^{-}  -\frac{1}{k}\rho
v_{h}\mathcal{G}-\mathbf{v}\cdot\mathbf{B}\,\mathcal{F}\right)  d^{3}r\,.
\end{align}
The  first variation of  the latter expression is given by 
\begin{align}
\delta\mathfrak{F} &  =%
{\displaystyle\int_{V}}
\left[  \left(  \rho\mathbf{v}_{\bot}-\mathbf{B}_{\bot}\,\mathcal{F}\right)\cdot
\delta\mathbf{v}_{\bot}+\left(  \rho v_{h}-B_{h}\,\mathcal{F}-\frac{1}{k}%
\rho\mathcal{G}\right)  \delta v_{h}\right.  \nonumber\\
&  +\left(  \frac{v^{2}}{2}+U+\frac{p}{\rho}-\mathcal{J}-\frac{1}{k}%
v_{h}\mathcal{G}\right)  \delta\rho+\left(  \frac{B_{h}}{4\pi}-k\mathcal{H}%
-v_{h}\,\mathcal{F}\right)  \delta B_{h}\nonumber\\
&  +\left(  -\nabla\cdot\left(  \frac{k^{2}}{4\pi}\mathbf{\nabla}\psi\right)
+\rho T\mathcal{S}^{\prime}-\rho\mathcal{J}^{\prime}-kB_{h}\mathcal{H}%
^{\prime}- k^{4}\left[  l\right]  \sin2\alpha\,  \mathcal{H}%
\right.  \nonumber\\
&  \left.  \left.  -\frac{1}{k}\rho v_{h}\mathcal{G}^{\prime}-\mathbf{v\cdot
B}\,\mathcal{F}^{\prime}+\nabla\cdot\left(  \mathcal{F}k\mathbf{h}%
\times\mathbf{v}_{\bot}\right)  \right)  \delta\psi\right]  d^{3}r\,.
\label{deF}
\end{align}
Here we have integrated by parts and neglected surface terms consistent with assumed boundary conditions.
Symmetric equilibria thus satisfy the set of equations%
\begin{align}
\rho\mathbf{v}_{\bot}-\mathbf{B}_{\bot}\,\mathcal{F} &  =0,\label{eq:eq1}\\
\rho v_{h}-B_{h}\,\mathcal{F}-\frac{1}{k}\rho\mathcal{G} &  =0,\label{eq:eq2}%
\\
\frac{v^{2}}{2}+U+\frac{p}{\rho}-\mathcal{J}-\frac{1}{k}v_{h}\mathcal{G} &
=0,\label{eq:eq3}\\
\frac{B_{h}}{4\pi}-k\mathcal{H}-v_{h}\mathcal{F} &  =0,\label{eq:eq4}\\
-\mathbf{\nabla}\cdot\left(  \frac{k^{2}}{4\pi}\mathbf{\nabla}\psi\right)
+\rho T\mathcal{S}^{\prime}-\rho\mathcal{J}^{\prime}-kB_{h}\mathcal{H}%
^{\prime}-   k^{4}\left[  l\right]  \sin2\alpha\,   \mathcal{H} \qquad &
\nonumber\\
-\frac{1}{k}\rho v_{h}\mathcal{G}^{\prime}-\mathbf{v\cdot B}\,\mathcal{F}%
^{\prime}+\nabla\cdot\left(  \mathcal{F}k\mathbf{h}\times\mathbf{v}_{\bot
}\right)   &  =0.\label{eq:eq5}%
\end{align}
Equations (\ref{eq:eq2}) and (\ref{eq:eq4}) can be combined to obtain%
\begin{equation}
v_{h}   =\left(  4\pi k\mathcal{H}\frac{\mathcal{F}}{\rho}+\frac{\mathcal{G}%
}{k}\right)  \left(  1-\mathrm{M}^{2}\right)  ^{-1} \qquad {\rm and} \qquad
B_{h}   =\left(  4\pi k\mathcal{H}+4\pi\mathcal{F}\frac{\mathcal{G}}%
{k}\right)  \left(  1-\mathrm{M}^{2}\right)  ^{-1},
\label{eq:Bh}
\end{equation}
which are two explicit relationships for  $v_{h}$ and $B_{h}$  that
make it possible  to express these two variables in terms of the flux function, the
cylindrical radius (which appears in $k$),  and the plasma density. The
dimensionless parameter $\mathrm{M}^{2}=4\pi\mathcal{F}^{2}/\rho$\ that
appears in the first of   Eqs.~(\ref{eq:Bh}) is the square of the Alfv\'{e}n Mach number. Notice that
on  Alfv\'en surfaces, i.e. points where $\mathrm{M}=1$, the regularity condition (see e.g. Ref.~\onlinecite{lovelace})
\begin{equation}
4\pi k\mathcal{H}\frac{\mathcal{F}}{\rho} + \frac{\mathcal{G}}{k}  = 0 \qquad \Leftrightarrow\qquad
4\pi k\mathcal{H}+4\pi\mathcal{F}\frac{\mathcal{G}}%
{k} = 0,
\label{eq:regcondbis}
\end{equation}
needs to be satisfied. In general, given the flux functions $\mathcal{F}$, $\mathcal{G}$,  and $\mathcal{H}$ and the boundary conditions, we can only check a posteriori whether the regularity condition is satisfied or not (of course, compatibility of the flux functions can be assessed a priori; for example,  if $\mathcal{F}>0$ and $\mathcal{G}>0$,  then $\mathcal{H}<0$).

Equation (\ref{eq:eq3}) gives a relationship  between the plasma density,  the
magnetic flux function  and its gradient, and $k$, %
\begin{equation}
\frac{k^{2}}{2}\left\vert \mathbf{\nabla}\psi\right\vert ^{2}\left(
\frac{\mathcal{F}}{\rho}\right)^{2} +\frac{v_{h}^{2}}{2}+  U+{p}/{\rho} - \frac{v_{h}}{k} \, \mathcal{G}=\mathcal{J}\,,%
\label{bern}
\end{equation}
where $U+{p}/{\rho}$ is the enthalpy. Equation (\ref{bern}),  a generalization of the  Bernoulli equation of hydrodynamics, and can be viewed as an equation for the density $\rho$ given $\psi$, making use of the second of Eqs.~(\ref{eq:Bh}) and a  particular choice of the Casimir functions $\mathcal{F}, \mathcal{G}, \mathcal{H}$,  and $\mathcal{J}$; however, in general it is not possible to obtain an explicit form for $\rho$.

The first term in Eq.~(\ref{eq:eq5}) can be rewritten in terms of the variables $r$ and $u$ as
\begin{equation}
\mathbf{\nabla}\cdot\left(  \frac{k^{2}}{4\pi}\mathbf{\nabla}\psi\right)
=\frac{1}{4\pi r^{2}}\left[  \frac{\partial^{2}\psi}{\partial u^{2}}%
+r\frac{\partial}{\partial r}\left(  rk^{2}\frac{\partial\psi}{\partial
r}\right)  \right]
\end{equation}
which corresponds to the differential operator of the so-called JOKF equation  \cite{JOKF, kadom,BOGO1}.
Moreover,  by using
Eq.~(\ref{eq:eq1}),  the last two terms of Eq.~(\ref{eq:eq5}) can be manipulated to obtain the following expressions:%
\begin{equation}
\mathbf{v\cdot B}\,\mathcal{F}^{\prime}=v_{h}B_{h}\mathcal{F}^{\prime}%
+k^{2}\left\vert \mathbf{\nabla}\psi\right\vert ^{2}\frac{\mathcal{FF}%
^{\prime}}{\rho}
\qquad {\rm and}\qquad
\mathbf{\nabla}\cdot\left(  \mathcal{F}k\mathbf{h}\times\mathbf{v}_{\bot
}\right)  =\mathbf{\nabla}\cdot\left(  \frac{k^{2}\mathcal{F}^{2}}{\rho
}\mathbf{\nabla}\psi\right)\,.
\end{equation}
Then,  Eq.~(\ref{eq:eq5})  becomes%
\begin{align}
\mathbf{\nabla}\cdot\left[  \left(  1-\mathrm{M}^{2}\right)  \frac{k^{2}}%
{4\pi}\mathbf{\nabla}\psi\right]  +k^{2}\left\vert \mathbf{\nabla}%
\psi\right\vert ^{2}\frac{\mathcal{FF}^{\prime}}{\rho}  & =\rho\left(
T\mathcal{S}^{\prime}-\mathcal{J}^{\prime}-v_{h}\frac{\mathcal{G}^{\prime}}%
{k}\right)  \nonumber\\
& -B_{h}\left(  k\mathcal{H}^{\prime}+v_{h}\mathcal{F}^{\prime}\right)
-\left(  k^{4}\left[  l\right]  \sin2\alpha\right)  \mathcal{H}\,,%
\label{genJOKF}
\end{align}
which is a  generalization of the JOKF equation that includes flow.

The  above equations were previously presented in Ref.~\onlinecite{tsinganos2} and various special solutions were obtained by  several authors \cite{villata, villata2, villata3, palumbo, zanna, palumbo2, tasso}.  However, the general variational principle $\delta \mathfrak{F}=0$ for helical equilibria with flow appears to be new, as well as reduced variational principles that we subsequently obtain by eliminating the algebraic constraints.

Upon choosing $k=1/r$ and $\alpha=0$, Eq.~(\ref{genJOKF})  reduces to  the azimuthally symmetric case and one obtains  the generalized Grad Shafranov equation with flow discussed in Ref.~\onlinecite{amp0}.  Similarly, upon choosing $k=1$ and $\alpha=\pi/2$, this  equation reduces to the translationally symmetric case  discussed in Ref.~\onlinecite{goedbloed1997}.
As discussed in Refs.~\onlinecite{goedbloed} and \onlinecite{lovelace}, the equation for the generalized
equilibria is hyperbolic for $\mathrm{M}_{c}^{2}\leq \mathrm{M}^{2}\leq \mathrm{M}_{s}^{2}$ and for $%
\mathrm{M}^{2}\geq \mathrm{M}_{f}^{2}$, where $\mathrm{M}_{c}^{2}\equiv \gamma p/\left( \gamma
p+B^{2}/4\pi \right) $ is the square Alfv\'{e}n Mach number corresponding to
the ``cusp velocity'' and
\begin{equation}
\mathrm{M}_{f,s}^{2}\equiv \frac{{4\pi \gamma p+B^{2}}}{{2B_{\perp}^{2}}}\left\{ 1\pm %
\left[ 1-\frac{{16\pi \gamma pB_{\perp}^{2}}}{{(4\pi \gamma p+B^{2})^{2}}}\right]
^{1/2}\right\}  \label{eq:mach_fast&slow}
\end{equation}%
is  that  relative to the fast  and slow
magnetosonic velocities,\ respectively $\mathrm{M}_{f}^{2}$\ and $\mathrm{M}_{s}^{2}$.

The variational principle of (\ref{deF}) can be reduced in several steps by `back-substituting' various algebraic relations.  First, by substituting the expression for the perpendicular velocity given by Eq.~(\ref{eq:eq1}) into the functional $\mathfrak{F}$ we  obtain a variational
principle that depends on the reduced set of independent variables, $\psi,\rho,v_{h}$, and $B_{h}$, viz.
\begin{align}
\mathfrak{F}\left[  \psi,\rho,v_{h},B_{h}\right]   &  =%
{\displaystyle\int_{V}}
\left(  \frac{\rho v_{h}^{2}}{2}+\rho U+\left(  1-\mathrm{M}^{2}\right)
\frac{k^{2}\left\vert \nabla\psi\right\vert ^{2}}{8\pi}+\frac{B_{h}^{2}}{8\pi
}  - \rho\mathcal{J}  \right.\nonumber\\
&  \left. \hspace{1.0 cm}  -kB_{h}\mathcal{H}-  k^{4}\left[  l\right]
\sin2\alpha\, \mathcal{H}^{-}  -\frac{1}{k}\rho
v_{h}\mathcal{G}-v_{h}B_{h}\,\mathcal{F}\right)   d^{3}r\,.
\end{align}
Similarly, we can reduce further by using Eq.~(\ref{eq:eq4}) to eliminate $B_{h}$,  yielding,
\begin{align}
\mathfrak{F}\left[  \psi,\rho,v_{h}\right]   &  =%
{\displaystyle\int_{V}}
\left(  \frac{\rho v_{h}^{2}}{2}+\rho U+\left(  1-\mathrm{M}^{2}\right)
\frac{k^{2}\left\vert \nabla\psi\right\vert ^{2}}{8\pi}-\frac{1}{8\pi}\left(
4\pi k\mathcal{H}+4\pi v_{h}\,\mathcal{F}\right)  ^{2}\right.  -    \rho\mathcal{J}  \nonumber\\
&  \left. \hspace{1.0 cm}-  k^{4}\left[  l\right]  \sin2\alpha\, \mathcal{H}^{-}  -\frac{1}{k}\rho v_{h}\mathcal{G}\right)
d^{3}r\,.
\label{vpvhrhopsi}
\end{align}
Next, we can use the first expression of (\ref{eq:Bh}) to eliminate the
dependence on $v_{h}$, obtaining the functional
\begin{align}
\mathfrak{F}\left[  \psi,\rho\right]   &  =%
{\displaystyle\int_{V}}
\left[  \rho U+\left(  1-\mathrm{M}^{2}\right)  \frac{k^{2}\left\vert
\nabla\psi\right\vert ^{2}}{8\pi}-\rho\mathcal{J}-    k^{4}\left[
l\right]  \sin2\alpha\, \mathcal{H}^{-} \right.
\nonumber\\
&  \left. \hspace{2.0 cm}  -\left(  \frac{\rho\mathcal{G}^{2}}{2k^{2}}+2\pi k^{2}%
\mathcal{H}^{2}+4\pi\mathcal{HGF}\right)  \left(  1-\mathrm{M}^{2}\right)
^{-1}\right]  d^{3}r\,.
\label{vprhopsi}
\end{align}%
One could attempt to reduce further, but because of the form of (\ref{bern}),  the density cannot be explicitly  eliminated without making further assumptions.  However, the density can be viewed as an implicit  functional of $\psi$ through  (\ref{bern}).  Thus, in a sense,  we have a minimal variational principle in terms of the variable $\psi$ alone.

Although the variational principle of Eq.~(\ref{vprhopsi}) is minimal,  it may not be the most efficacious to use.  Observe, the last substitution  introduced a potential
singularity at  $\mathrm{M}=1$.    If we seek  extrema of  (\ref{vprhopsi}) by considering a sequence of  $L^{2}$ functions,  the principle (\ref{vprhopsi}) in general leads to  singularities on $\mathrm{M}=1$.  However,  if we expand $v_h$, $\rho$,  and $\psi$ and insert into the variational principle (\ref{vpvhrhopsi}), the quantity $v_{h}$ will always be regular and this also follows for the integrand.  Nevertheless, the principle of (\ref{vprhopsi}) may be useful.  For example, suppose $M$ depends only on $\psi$, which is the case for incompressible equilibria (cf. Ref.~\onlinecite{tasso}).   Then, the term
$\mathcal{E}_p:=   \int_V \left(  1-\mathrm{M}^{2}\right)  {k^{2}\left\vert
\nabla\psi\right\vert ^{2}}\, d^3r/{8\pi}$
can be simplified by a simple variable change from $\psi$ to a new variable $\chi$.  If we suppose $\psi=\Psi(\chi)$,  substitute into $\mathcal{E}_p$, and  set $\Psi'^2 \left(  1-\mathrm{M}^{2}\right)=1$ we obtain $\mathcal{E}_p = \int_V   {k^{2}\left\vert
\nabla\chi\right\vert ^{2}} \, d^3r/{8\pi}$.  Therefore,  the transformation
\begin{equation}
\chi=\int^{\psi} \sqrt{1-M^2(\psi')}\,  d\psi'
\label{mytrans}
\end{equation}
eliminates the $|\nabla \psi|$ term from (\ref{genJOKF}) and yields an equation in terms of $\chi$ that is identical to that without `poloidal' flow.   Thus one can use (\ref{mytrans}) to map equilibria without flow into to those with flow profiles determined by $M(\psi)$.  This transformation was first noted in Ref.~\onlinecite{morrison86} for two-dimensional axisymmetric equilibria and generalized, including the helical case,  in Ref.~\onlinecite{tasso}.

\section{Summary and Discussion}
\label{conclu}

In this paper we have written the noncanonical Hamiltonian structure of MHD in a general form that includes translational, azimuthal, and helical symmetry.   From the noncanonical Poisson bracket we obtained  Casimir invariants for all symmetries, including a new ones that did not appear in Ref.~\onlinecite{amp0}.  From these invariants we constructed variational principles for equilibria, including helical symmetry, and showed   how to reduce these variational principles to fewer numbers of variables.    A general equilibrium equation that includes general flow was presented.

The variational principles we obtained are useful for constructing solutions by the direct method of the calculus of variations  \cite{CH}.  One can insert sequences of functions and reduce the extremization to the solution of algebraic equations.  Approximate solutions for the case of axisymmetric  and fully 3D equilibria have been obtained in this way in Refs.~\onlinecite{betan,VMEC1,VMEC2,Garab1,Garab2}.  Similarly,  axisymmetric equilibria with flow have been obtained  for application to laboratory and astrophysical plasmas   \cite{goedbloed_evp} and plasma thrusters  \cite{ap1,ap2}.    Likewise, the variational principle of  (\ref{vprhopsi})  can be used to construct helical equilibria with and without flow that are of importance for both laboratory and naturally occurring plasmas.   We list several possibilities.

First,  the plasma thruster problem treated in Refs.~\onlinecite{ap1,ap2}  can be generalized to include the helical structures that have been observed to arise from the saturation of kink modes  \cite{zuin,zuin1}.   Ascertaining the nature of these structures is important for determining the effectiveness of these thrusters.   This will be the subject of a future publication.

Another potential application would be to analyze  helical structures called `snakes'  that were detected in the JET experiment at Culham  \cite{wesson}.  These structures,  detected by  soft X-ray emission, are formed by  local plasma cooling  caused by the ablation of  a pellet  injected  into the tokamak.   They have been  interpreted  as a persistent local modification along a closed magnetic field line of the global  toroidal axisymmetric  equilibrium.   This structure  in the plasma and its persistence  might be described as a  helical static equilibrium along the closed magnetic flux tube crossed by the pellet, and thus would be accessible by our variational principle.

Helical configurations that appear in Reverse Field  Pinch Configurations, the so-called Quasi Single Helicity states (e.g. Ref.~\onlinecite{zuin})  presents another application.  These states result from plasma self-organization,  where a dominant mode tends to suppress modes with different helicity,  and have reduced magnetic  turbulence  and better energy confinement. Since all these helical states  have a large aspect ratio, toroidal curvature effects may be neglected to first order and  their equilibrium configuration can be described by our variational principles.   Helical structures (flux ropes)  are also  found  to arise in numerical simulations of three-dimensional magnetic reconnection processes \cite{Daugh}.
\\ Similarly, helical equilibria can be used to model straight (large aspect ratio) stellarator configurations (e.g. Ref.~\onlinecite{booze}).    The Helically Symmetric Experiment at  Madison Wisconsin  \cite{Wis}  has a Quasi-Helically Symmetric magnetic field structure and  thus   avoids the consequences of the lack of symmetry in the magnetic fields in conventional stellarators that  results in large deviations of particle orbits from magnetic surfaces and direct loss orbits.

Helical equilibria  are of special importance for space configurations where they arise naturally as the result of the plasma  streaming and kinking. In this context the problem of the existence of ``regular''  helical equilibria was addressed  in the context of a long lasting dispute about the so-called Parker theorem that, loosely formulated, implies that in the absence of translational invariance, current layers (tangential discontinuities) must form in  MHD static equilibria.  This issue appears to have been settled definitively in Refs.~\onlinecite{BOGO1,BOGO2}, by the  explicit  construction of globally regular helical solutions for helical equilibria.   These solutions are of  mathematical interest since they show that helical equilibrium solutions can be found as continuous deformations of cylindrically symmetric equilibria.  At  the same time they provide useful models of  plasma jets in space.  The extension from static to stationary helical equilibria (i.e.\  equilibria with flow) is of  major interest for the description of plasma jets in space. In this case exact solutions of our generalized JOKF equation  (\ref{eq:eq5})  can be searched for by means of our  reduced variational principle (\ref{vprhopsi}), in a manner similar to that used to obtain the  axisymmetric thruster equilibria of Refs.~\onlinecite{ap1,ap2}.

Obtaining equilibria that are extrema of the variational principle (\ref{vprhopsi}) allows us to consider their stability by effecting the second variation.  We will consider a variety of such energy stability calculations for a variety of equilibrium states in Ref.~\onlinecite{amp2}.

\section*{Acknowledgment}
\noindent  PJM was supported by U.S. Dept.\ of Energy Contract \# DE-FG05-80ET-53088.


\begin{thebibliography}{99}                                                                                               %

\bibitem{freidberg}
J. P. Freidberg,  {\em Ideal Magnetohydrodynamics (Modern Perspectives in Energy)} (Springer, Berlin, 1987).


\bibitem{goedbloed} J. P. Goedbloed and  S. Poedts,    {\em Principles of Magnetohydrodynamics: With Applications to Laboratory and Astrophysical Plasmas} (Cambridge University Press, Cambridge, 2004).

\bibitem {woltjer1a}L. Woltjer,
\textit{Proc. Natl. Acad. Sci. USA} \textbf{44}, 833--41 (1958); \textit{Proc. Natl. Acad. Sci. USA} \textbf{45}, 769--71 (1959).

\bibitem {woltjer2a}L. Woltjer,
\textit{Astrophys J.} \textbf{130}, 400--4 (1959); ibid, 405--13.

%
%
%

\bibitem {taylor1}J. B. Taylor,
\textit{Phys. Rev. Lett.} \textbf{33}, 1139--41 (1974).

\bibitem {taylor2}J. B. Taylor,
\textit{Rev. Mod. Phys.} \textbf{58}, 741--63 (1986).

\bibitem {hameiri1}E. Hameiri,
\textit{Phys. Plasmas} \textbf{5}, 3270--81 (1998).

\bibitem {goedbloed_evp}J. P. Goedbloed,
\textit{Phys. Plasmas} \textbf{11}, L81--3 (2004).

\bibitem {ap1}T. Andreussi and F. Pegoraro,
\textit{Phys. Plasmas} \textbf{15}, 092108-1--7 (2008)

\bibitem{04Ilin} K. I. Ilin and V. A. Vladimirov, 
\textit{Phys. Plasmas} \textbf{11}, 3586-3593 (2004).

\bibitem{Bernstein1958b}
I. B. Bernstein, E. A. Frieman, M. D. Kruskal, and R. M. Kulsrud, 
\textit{Proc. R. Soc. Lond. A} \textbf{244},  17  (1958).

\bibitem{hain}
K. Hain, R. L\"{u}st, and A. Schl\"{u}ter,  
\textit{Z. Naturforsch.} \textbf{12a},  833  (1957).

\bibitem{Frieman1960}
E. A. Frieman and M. Rotenberg, 
\textit{Rev. Mod. Phys.} \textbf{32},  898  (1960).

\bibitem {newcomb}W. A. Newcomb,
\textit{Nucl. Fusion: Supplement} pt \textbf{2}, 451--63 (1962).

\bibitem {morrison-greene}P. J. Morrison and J. M. Greene,
\textit{Phys. Rev. Lett.} \textbf{45}, 790--4 (1980).

\bibitem {morrison-greene-add}P. J. Morrison and J. M. Greene,
\textit{Phys. Rev. Lett.} \textbf{48}, 569 (1982).

\bibitem {morrison1}P. J. Morrison,
\textit{AIP Conf. Proc.} \textbf{88}, 13--45 (1982).


\bibitem{lagrange}J. L. Lagrange,  \emph{M\'ecanique Analytique},  English Title:  Analytical Mechanics,  translated and edited by A.~Boissonnade and V.N.~Vagliente (Kluwer Academic, Imprint Dordrecht,  Boston, Mass. 1997).

\bibitem{koslov}V. I. Arnold, V. V. Kozlov, and A. I. Neishtadt,  {\em
Dynamics III: Mathematical Aspects of Classical and Celestial Mechanics} (Springer, Berlin, 1990).

\bibitem{dirichlet} P. G. L. Dirichlet, Crelle \textbf{32}, 3 (1846), (Desaint, Paris).  See Note II of   Lagrange, J. L.,  {\it M\'{e}canique Analytique}, Quatri\`{e}me \'Edition,  Tome Premier, (Gauthier-Villars et Fils, Paris, 1888) p. 457.



\bibitem{Morrison1998}P. J. Morrison, 
\textit{Rev. Mod. Phys.} \textbf{70},  467  (1998).

\bibitem{amp2}T. Andreussi,   P. J. Morrison,  and F. Pegoraro, 
under preparation (2012).

\bibitem{amp0}T. Andreussi,   P. J. Morrison,  and F. Pegoraro,
\textit{Plasma Phys. Contr. Fusion} \textbf{52}, 055001--1--22 (2010).

\bibitem{gudunov}S. K. Gudunov, 
\textit{Dokl. Akad. Nauk.} \textbf{139}, 521 (1961).

\bibitem {padhye1}N. Padhye and P. J. Morrison,
\textit{Phys. Lett.} A \textbf{219}, 287--92 (1996).

\bibitem {padhye2}N. Padhye and P. J. Morrison,
\textit{Plasma Phys. Rep.} \textbf{22}, 869--77 (1996).

\bibitem {lovelace}R. V. E. Lovelace, C. Mehanian, C. M. Mobarry, and M. E. Sulkanen,
\textit{Astrophys. J. Suppl. Ser.} \textbf{62}, 1--37 (1986).

\bibitem{JOKF} J. L.  Johnson,  C. R. Oberman,   R. M.  Kulsrud,  and  E. A. Frieman,
\textit{Phys. Fluids} \textbf{1}, 281--96 (1958).

\bibitem{kadom} B. B. Kadomtsev, 
\textit{Soviet Physics JETP} \textbf{10}, 962--963 (1960).

\bibitem{BOGO1}  O. I. Bogoyavlenskij,  
\textit{Lett. Math.  Phys.} \textbf{51},  235--247 (2000).

\bibitem {tsinganos2}K. C. Tsinganos,
\textit{Astrophys. J.} \textbf{259}, 820--31 (1982).

\bibitem{palumbo}L. J. Palumbo and A. M. Platzeck, 
\textit{Astrophys. J.} \textbf{416}, 656 (1993).

\bibitem{villata}M. Villata and K. Tsinganos, 
\textit{Phys. Fluids B} \textbf{5}, 2153 (1993).

\bibitem{villata2}M. Villata and A. Ferrari, 
\textit{Astron. Astrophys.} \textbf{284}, 663 (1994).

\bibitem{villata3}M. Villata and A. Ferrari, 
\textit{Phys. Plasmas} \textbf{1}, 2200 (1994).

\bibitem{zanna}L. Del Zanna and C. Chiuderi, 
\textit{Astron. Astrophys.} \textbf{310}, 341 (1996).

\bibitem{palumbo2}L. J. Palumbo and A. M. Platzeck, 
\textit{J. Plasma Phys.} \textbf{60}, 449 (1998).

\bibitem{tasso}G. N. Throumoulopoulos and H. Tasso, 
\textit{J. Plasma Phys.} \textbf{62}, 449 (1999).

\bibitem{goedbloed1997}J. P. Goedbloed and  A. Lifschitz, 
\textit{Phys. Plasmas} \textbf{4}, 3544 (1997).

\bibitem{morrison86} P. J. Morrison, 
\textit{Bull. Am. Phys. Soc.} \textbf{31}, 1609  (1986).

\bibitem{CH}R. Courant and D. Hilbert, \emph{Methods of Mathematical Physics I} (New York: Wiley
  Interscience, 1953).

\bibitem{betan}O. Betancourt and P. Garabedian,  
\textit{Proc. Nat. Acad. Sci. USA} \textbf{73},  984 (1976).

\bibitem{VMEC1}S.P. Hirshman and J.C. Whitson,
\textit{Phys. Fluids} \textbf{26}, 3554 (1983).

\bibitem{VMEC2}J.D. Hanson,  S. P. Hirshman, S. F. Knowlton,  S. F. Knowlton, L. L. Lao, E.  A. Lazarus, and J. M. Shields, 
\textit{Nucl. Fusion} \textbf{49}, 075031 (2009).

\bibitem{Garab1}P. R. Garabedian,
\textit{Proc. Natl. Acad. Sci. USA} \textbf{99}, 10257 (2002).

 \bibitem{Garab2}P. R. Garabedian,
\textit{Proc. Natl. Acad. Sci. USA} \textbf{100}, 13741 (2003).


\bibitem {ap2}T. Andreussi and F. Pegoraro,
\textit{Phys. Plasmas} \textbf{17}, 063507-1--11 (2010).

\bibitem{zuin}M.  Zuin, R. Cavazzana, E. Martines,  G. Serianni,  V. Antoni, M. Bagatin, M. Andrenucci, F. Paganucci, and P. Rossetti,  
\textit{Phys. Rev. Lett.} \textbf{92}, 5003 (2004).

\bibitem{zuin1}W. F. Bergerson, F. Auriemma,  B. E. Chapman,  W. X. Ding,  P. Zanca,  D. L. Brower,  P. Innocente,  L. Lin, R. Lorenzini,  E. Martines, B. Momo, J. S. Sarff, and D. Terranova,
\textit{Phys. Rev. Lett.} \textbf{107}, 5003 (2011).

\bibitem{wesson}J. A. Wesson,  
\textit{Plasma Phys. Control. Fusion} \textbf{37},  A337-A346 (1995).

\bibitem{Daugh}
W. Daughton, V. Roytershteyn, H. Karimabadi, L. Yin, B. J. Albright, B. Bergen, and K. J. Bowers, 
\textit{Nature Phys.} \textbf{7}, 539 (2011).

\bibitem{booze}L. P.  Ku and A. H. Boozer, 
\textit{Nucl. Fusion} \textbf{51}, 013004 (2011).
  	
\bibitem{Wis}J. M. Canik, D. T. Anderson, F. S. B. Anderson, K. M. Likin, J. N. Talmadge, and K. Zhai,
\textit{Phys. Rev. Lett.} \textbf{98}, 085002 (2007).

\bibitem{BOGO2}O. I. Bogoyavlenskij,  
\textit{Phys. Rev.} E \textbf{62},  8616--27 (2000).


\bibitem{hanson2}J. D. Hanson and S. P. Hirshman,
\textit{Phys. Plasmas} \textbf{9}, 4410 (2002).

\end{thebibliography}
\end{document}